\documentstyle[aps,prb,multicol]{revtex}
\begin{document}

\title{Penetration Depth and the Conductivity Sum Rule for a
       Model With Incoherent c-axis Coupling}
\author{Eugene H. Kim}
\address{Department of Physics, University of California \\
         Santa Barbara, California 93106-9530}
\maketitle
 
\begin{abstract}
The conductivity sum rule for a one-band hopping model relates
the integrated spectral weight of the real part of the conductivity
to the average kinetic energy.  For such a model, the superconducting
penetration depth is therefore dependent upon both the change in the 
conductivity spectral weight and the change in kinetic energy between
the normal and superconducting states.  Here we examine the consequences
of this for the c-axis penetration depth of a layered system in which
the charge transfer perpendicular to the layers (along the c-axis)
is mediated by interlayer impurity scattering.
\end{abstract}

\vspace{.3in}
\begin{multicols}{2}
The nature of the frequency and temperature dependent c-axis
conductivity, $ \sigma_{1c}(\omega,T) $, in the cuprate
superconductors remains controversial, but for a number of these
materials it appears to be weak and incoherent\cite{new1}.  Recently,
a simple model\cite{old1,old2,old3} consisting of layers with BCS
quasiparticles which have a $ d_{x^2 - y^2} $ gap and an interlayer
coupling mediated by impurity scattering was used to calculate $
\sigma_{1c}(\omega,T) $.  For this model, the conductivity sum rule
relates the integrated spectral weight under $ \sigma_{1c}(\omega,T) $
to the average kinetic energy per unit cell in the
c-direction\cite{old4}.  For such a model, the superconducting
penetration depth is dependent upon both the change in the
conductivity spectral weight and the change in the kinetic energy.
Here we examine this and discuss its consequences.

We consider a Hamiltonian of the form
\begin{equation}
    H = H_{ab} + H_c
\end{equation}
where $ H_{ab} $ describes the intralayer dynamics and $ H_c $ is the 
interlayer coupling
\begin{equation}
    H_c = \sum_{l,s}  V_l  ( c_{l+z,s}^{\dagger}c_{l,s} +
                            c_{l,s}^{\dagger}c_{l+z,s} )
\end{equation}
Here $ V_l $ is a random potential due to impurity scattering between
layers.  We assume that $ H_{ab} $ describes quasiparticles with 
energy $ \varepsilon_p $ in the normal state and BCS quasiparticles
with dispersion $ E_p = \sqrt{\varepsilon_p^2 + \Delta_p^2 } $ in the
superconducting state with $ \Delta_p = \Delta_0\cos2\phi_p $, a 
 $ d_{x^2 - y^2} $ gap.

For this model, the c-axis conductivity sum rule has the form\cite{old4}
\begin{equation} 
    \frac{2}{\pi e^{2}d^{2} } \int_{0}^{\infty} \sigma_{1c}(\omega) d\omega = 
    - <K_c> 
\end{equation}
where $ d $ is the interlayer spacing, and $ <K_c> $ is the c-axis
kinetic energy per unit volume 
\begin{equation}
    <K_c> = \frac{<H_c>}{V}
\end{equation}
If the change in $ <K_c> $ between the normal and superconducting states
is negligible, one has the usual relationship between the loss in the
$ (\omega > 0) $ spectral weight of the conductivity in the 
superconducting state relative to the normal state and the c-axis
penetration depth, $ \lambda_c $\cite{old5,old6}
\begin{equation}
    \frac{c^{2}}{4 \pi \lambda_{c}^{2} } =
    \frac{2}{\pi } \int_{0^{+}}^{\infty} 
    d\omega ( \sigma_{1c}^{N}(\omega) - \sigma_{1c}^{S}(\omega) )
\end{equation}
Here $ \sigma_{1c}^N $ and $ \sigma_{1c}^S $ are the normal and
superconducting c-axis conductivities, respectively.
However, when the c-axis tunneling process is incoherent and the gap has a 
strong momentum dependence, the change in $<K_c>$ between the superconducting
and normal states becomes important.  Then Eqn.(5) is modified to
\[
   \frac{c^{2}}{4 \pi \lambda_{c}^{2} } =
    \frac{2}{\pi } \int_{0^{+}}^{\infty}
    d\omega ( \sigma_{1c}^{N}(\omega) - \sigma_{1c}^{S}(\omega)  )
\]
\vspace{-.27in}
\begin{equation}
   - \, e^2 d^2 ( <K_c>^{S} - <K_c>^{N} )
\end{equation}
For the case of a $ d_{x^{2}-y^{2}} $ superconductor, if the tunneling
process is diffuse, the Josephson coupling between the layers
vanishes\cite{old1,old2,old3} and $\lambda_c$ is infinite.  In this
case, $\sigma_{1c} (\omega)$ is still suppressed when the gap is
opened (see Fig. 2 of Ref.~\onlinecite{old3}) but the change in the
kinetic energy in Eqn.(6) cancels the change in the spectral weight,
leading to an infinite $ \lambda_c $.  If the incoherent tunneling
process is anisotropic, there will only be a partial cancellation,
leading to a larger $ \lambda_c $ than one would find using Eqn.(5).
Here, we examine this effect for an impurity scattering model of the
interlayer transport.

Taking $ V_l $ to be weak, the first non-vanishing contribution to $
<K_c> $, after averaging over impurities\cite{old3,old9}, is
\[
   \frac{4n^c_{imp}}{N^2_{ab}} \sum_{k,p} \overline{|V_{pk}|^2} 
   T \sum_{n} 
   \frac{(i \omega_n + \epsilon_p)(i \omega_n + \epsilon_k)}
        { [(i \omega_n)^{2} - E_{p}^{2} ][(i \omega_n)^{2} - E_{k}^{2} ] }  
\]
\vspace{-.27in}
\begin{equation}
   -\,  \frac{4n^c_{imp}}{N^2_{ab}} \sum_{k,p} \overline{|V_{pk}|^2}
    T \sum_{n} \frac{ \Delta_k \Delta_p }
   { [(i \omega_n)^{2} - E_{p}^{2} ][(i \omega_n)^{2} - E_{k}^{2} ] } 
\end{equation}
where $ n^c_{imp} $ is the impurity concentration which causes
c-axis transport, $ N_{ab} $ is the number of sites in the $ab$ plane,
$ \omega_n = (2n + 1)\pi T$, and we will take the impurity
potential to have the separable form
\begin{equation}
    \overline{|V_{pk}|^2} \; = \; \mid V_0 \mid ^{2} +
                                 \mid V_1 \mid ^{2} 
                                  \cos 2 \phi_k \cos 2 \phi_p
\end{equation}
Physically, the first term in Eqn.(7) is due to quasi-particle fluctuations
between the layers, while the second term is due to superconducting
pair fluctuations.

Setting $ \Delta_k = 0 $ in Eqn.(7) gives 
us $ <K_c>^{N} $.  Taking $ \Delta_k = \Delta_0 \cos2 \phi_k $   
gives $ <K_c>^{S} $ for a $ d_{x^2-y^2} $ superconductor. Thus we
find that 
\[
    <K_c>^{S}_{d_{x^2-y^2}} - <K_c>^N =
\]
\vspace{-.27in}
\small{
\[
     - 16 n^c_{imp} N^2(0)
    \, T \sum_n \mid V_0 \mid^2
      [ \frac{\omega_n^2}{ \Delta_0^2 + \omega_n^2 }
     {\bf K}^2 ( \frac{ \Delta_0}{ \sqrt{ \Delta_0^2 + \omega_n^2 } })
     - ( \frac{\pi}{2} )^2 ]
\]
\vspace{-.27in}
\[
    - 16 n^c_{imp} N^2(0) \, T \sum_n
    \{ \frac{ \mid V_1 \mid ^2 }
            { \Delta_0^2 ( \Delta_0^2 + \omega_n^2 ) }
    [ \omega_n^2 {\bf K} ( \frac{ \Delta_0 }
                         { \sqrt{ \Delta_0^2 + \omega_n^2 } } )
\]
\vspace{-.27in}
\begin{equation}
    - ( \Delta_0^2 + \omega_n^2)
      {\bf E} ( \frac{ \Delta_0 }{ \sqrt{ \Delta_0^2 + \omega_n^2 } } )
       ]^2 \}
\end{equation}
      }
\normalsize       
where $ N(0) $ is the bare single particle density of states, and
{\bf K} and {\bf E} are complete elliptic integrals of the first
and second kinds, respectively\cite{old8}. 
For $T \ll \Delta_0$ ,
\[
   <K_c>^S_{d_{x^2-y^2}} - <K_c>^N = 
\]
\vspace{-.27in}
\[
    \frac{8n^c_{imp} N^2(0)}{\pi} \Delta_0
    \left( 5.12 |V_0|^2 - 2.37 |V_1|^2 \right)
\]
\vspace{-.27in}
\begin{equation}
   +  {\cal O} [(\frac{T}{\Delta_0})^3
    \ln^2(\frac{T}{\Delta_0})]
\end{equation}

Then from Eqn.(6) we have
\[
    \frac{c^2}{4\pi \lambda_c^2} = \frac{2}{\pi}
    \int_{0^+}^{\infty} d\omega 
    ( \sigma_{1c}^N(\omega) - \sigma_{1c}^S(\omega) )
\]
\vspace{-.26in}
\begin{equation}
     - \, \frac{8 n^c_{imp} N^2(0)}{\pi} e^2 d^2 \Delta_0
             ( 5.12 \mid V_0 \mid^2 \, - \, 2.37 \mid V_1 \mid^2 )
\end{equation}
However, we know that when $ \mid V_1 \mid^2 = 0 $ there is no pair
transport and $ \lambda_c $ becomes infinite.  In this case, the
 $ \mid V_0 \mid^2 $ term gives the difference between the area under
 $ \sigma_{1c}^N(\omega,T) $ and $ \sigma_{1c}^S(\omega,T) $ for 
 $ \omega > 0 $ and there is no $ \delta $-function contribution 
at $ \omega = 0 $.
For $ \mid V_1 \mid^2 $ small but
non-vanishing, $ \lambda_c $ becomes finite but {\em larger} than one
would estimate from the missing spectral area
 $ \sigma_{1c}^N(\omega,T) - \sigma_{1c}^S(\omega,T) $ for
 $ \omega > 0 $.  If $ \mid V_1 \mid^2 $ increases sufficiently so that
 $ \mid V_1 \mid^2 = 2.16 \mid V_0 \mid^2 $
 then there is no change
between $ <K_c>^S $ and $ <K_c>^N $ and the correct $ \lambda_c $ 
is obtained from the familiar sum rule, Eqn.(5).

Eqn.(18) of Ref.[4] gives a prediction for the c-axis penetration depth for the 
model we considered here.  It is
\begin{equation}
  \frac{c^2}{4\pi\lambda_c^2} \simeq 4\pi e^2 d^2 n_{imp}^c N^2(0)
                                     |V_1|^2 \Delta_0 (.48)
\end{equation}
This is the result one would obtain if a direct magnetic measurement of the
penetration depth were made.  Our eqn.(11) also gives a prediction for the
c-axis penetration depth.  However, eqn.(11) is the penetration depth
inferred from a measurement of the conductivity.  Our
results show that for a momentum-dependent gap, the conductivity sum
rule must be applied with care to determine the penetration depth.  Our 
results show that, for a momentum-dependent gap,  there is a change in the
c-axis kinetic energy between
the normal and superconducting states; this change in kinetic energy
must be taken into account in order to correctly obtain the penetration
depth from the conductivity sum rule.   A nieve application of the 
conductivity sum rule ( eqn.(5) ) would imply a penetration depth which
is smaller or larger than what would be measured.   From a correct application
of the sum rule ( eqn.(6) ), the correct value of the penetration 
depth could be inferred.   Eqn.(11) and eqn.(12) give the same
value for the penetration depth.  However, eqn.(11) is what one would use
to infer the penetration depth from a measurement of the conductivity.

One can ask what has happened to the conductivity spectral
weight.
As Hirsch discussed \cite{old7}, spectral weight can be transferred to or
from higher bands which are not included in our simple interlayer hopping
model.  Note however, for $ \mid V_1 \mid^2 < 2.16\mid V_0 \mid^2 $
we have the opposite effect to that discussed by Hirsch for his model
of hole superconductivity.  That is, for the impurity model we have
considered here, when the system goes into the superconducting state,
if $ \mid V_1 \mid^2 < 2.16\mid V_0 \mid^2 $, spectral weight is 
transferred to higher bands and the true $ \lambda_c $ is larger
than one would obtain by simply determining the missing spectral weight
according to Eqn.(5).  Conversely, if 
 $\mid V_1 \mid^2 > 2.16\mid V_0 \mid^2 $ spectral weight is transferred 
down from higher bands and the true $ \lambda_c $ is actually smaller
than that given by Eqn.(5).

The author would like to thank D.~J.~Scalapino for suggesting this 
problem.  The author also thanks D.~J.~Scalapino, D.~Q.~Duffy, and 
C.~L.~Martin for
useful discussions.  This work was supported by NSF grant No. 
DMR-9527304.

\vspace{.3in}

\end{multicols}
\end{document}